\documentstyle[12pt]{article}

\input epsf

\begin{document}
\baselineskip=20.5pt

\def\beqra{\begin{eqnarray}} \def\eeqra{\end{eqnarray}}
\def\beqast{\begin{eqnarray*}}
\def\eeqast{\end{eqnarray*}}
\def\beq{\begin{equation}}      \def\eeq{\end{equation}}
\def\be{\begin{enumerate}}   \def\ee{\end{enumerate}}

\def\fnote#1#2{\begingroup\def\thefootnote{#1}\footnote{
#2}
\addtocounter
{footnote}{-1}\endgroup}

\def\itp#1#2{\hfill{NSF-ITP-{#1}-{#2}}}

\def\gam{\gamma}
\def\Gam{\Gamma}
\def\la{\lambda}
\def\eps{\epsilon}
\def\La{\Lambda}
\def\si{\sigma}
\def\Si{\Sigma}
\def\al{\alpha}
\def\Tha{\Theta}
\def\tha{\theta}
\def\vphi{\varphi}
\def\del{\delta}
\def\Del{\Delta}
\def\ab{\alpha\beta}
\def\om{\omega}
\def\Om{\Omega}
\def\mn{\mu\nu}
\def\mun{^{\mu}{}_{\nu}}
\def\kap{\kappa}
\def\rsi{\rho\sigma}
\def\beal{\beta\alpha}

\def\til{\tilde}
\def\rta{\rightarrow}
\def\eqv{\equiv}
\def\nab{\nabla}
\def\pa{\partial}
\def\sit{\tilde\sigma}
\def\ul{\underline}
\def\indt{\parindent2.5em}
\def\nd{\noindent}

\def\rsi{\rho\sigma}
\def\beal{\beta\alpha}

\def\caa{{\cal A}}
\def\cb{{\cal B}}
\def\cac{{\cal C}}
\def\cd{{\cal D}}
\def\ce{{\cal E}}
\def\cf{{\cal F}}
\def\cg{{\cal G}}
\def\cah{{\cal H}}
\def\ci{{\cal I}}
\def\cj{{\cal{J}}}
\def\ck{{\cal K}}
\def\cl{{\cal L}}
\def\cm{{\cal M}}
\def\cn{{\cal N}}
\def\cO{{\cal O}}
\def\cp{{\cal P}}
\def\car{{\cal R}}
\def\cs{{\cal S}}
\def\ct{{\cal{T}}}
\def\cu{{\ca{U}}}
\def\cv{{\cal{V}}}
\def\cw{{\cal{W}}}
\def\cx{{\cal{X}}}
\def\cy{{\cal{Y}}}
\def\cz{{\cal{Z}}}

\def\raisenot{\raise .5mm\hbox{/}}
\def\nota{\ \hbox{{$a$}\kern-.49em\hbox{/}}}
\def\notA{\hbox{{$A$}\kern-.54em\hbox{\raisenot}}}
\def\notb{\ \hbox{{$b$}\kern-.47em\hbox{/}}}
\def\notB{\ \hbox{{$B$}\kern-.60em\hbox{\raisenot}}}
\def\notc{\ \hbox{{$c$}\kern-.45em\hbox{/}}}
\def\notd{\ \hbox{{$d$}\kern-.53em\hbox{/}}}
\def\notbd{\ \hbox{{$D$}\kern-.61em\hbox{\raisenot}}} 
\def\note{\ \hbox{{$e$}\kern-.47em\hbox{/}}}
\def\notk{\ \hbox{{$k$}\kern-.51em\hbox{/}}}
\def\notp{\ \hbox{{$p$}\kern-.43em\hbox{/}}}
\def\notq{\ \hbox{{$q$}\kern-.47em\hbox{/}}}
\def\notW{\ \hbox{{$W$}\kern-.75em\hbox{\raisenot}}}
\def\notz{\ \hbox{{$Z$}\kern-.61em\hbox{\raisenot}}}
\def\notpa{\hbox{{$\partial$}\kern-.54em\hbox{\raisenot}}}

\def\fo{\hbox{{1}\kern-.25em\hbox{l}}}  
\def\rf#1{$^{#1}$}
\def\bx{\Box}
\def\tr{{\rm Tr}}
\def\rmtr{{\rm tr}}
\def\dgg{\dagger}

\def\lag{\langle}
\def\rag{\rangle}
\def\bmid{\big|}

\def\vlap{\overrightarrow{\La p}} 
\def\lrta{\longrightarrow}
\def\lrar{\raisebox{.8ex}{$\longrightarrow$}}
\def\rlarw{\longleftarrow\!\!\!\!\!\!\!\!\!\!\!\lrar}

\def\llra{\relbar\joinrel\longrightarrow}     
\def\mapright#1{\smash{\mathop{\llra}\limits_{#1}}}
\def\mapup#1{\smash{\mathop{\llra}\limits^{#1}}}
\def\asymptotic{{_{\stackrel{\displaystyle\longrightarrow}
{x\rightarrow\pm\infty}}\,\, }} 
\def\asymptext{\raisebox{.6ex}{${_{\stackrel{\displaystyle\longrightarrow}
{x\rightarrow\pm\infty}}\,\, }$}} 

\def\7#1#2{\mathop{\null#2}\limits^{#1}}   
\def\5#1#2{\mathop{\null#2}\limits_{#1}}   
\def\too#1{\stackrel{#1}{\to}}
\def\tooo#1{\stackrel{#1}{\longleftarrow}}
\def\nout{{\rm in \atop out}}

\def\one{\raisebox{.5ex}{1}}
\def\BM#1{\mbox{\boldmath{$#1$}}}

\def\ltsim{\matrix{<\cr\noalign{\vskip-7pt}\sim\cr}}
\def\gtsim{\matrix{>\cr\noalign{\vskip-7pt}\sim\cr}}
\def\haf{\frac{1}{2}}


\def\place#1#2#3{\vbox to0pt{\kern-\parskip\kern-7pt
                             \kern-#2truein\hbox{\kern#1truein #3}
                             \vss}\nointerlineskip}

\def\illustration #1 by #2 (#3){\vbox to #2{\hrule width #1
height 0pt
depth
0pt
                                       \vfill\special{illustration #3}}}

\def\scaledillustration #1 by #2 (#3 scaled #4){{\dimen0=#1
\dimen1=#2
           \divide\dimen0 by 1000 \multiply\dimen0 by #4
            \divide\dimen1 by 1000 \multiply\dimen1 by #4
            \illustration \dimen0 by \dimen1 (#3 scaled #4)}}

\def\ON{{\cal O}(N)}
\def\UN{{\cal U}(N)}
\def\bdPh{\mbox{\boldmath{$\dot{\!\Phi}$}}}
\def\bPh{\mbox{\boldmath{$\Phi$}}}
\def\bPhs{\bPh^2}
\def\sef{S_{eff}[\sigma,\pi]}
\def\sigx{\sigma(x)}
\def\pix{\pi(x)}
\def\bph{\mbox{\boldmath{$\phi$}}}
\def\bphs{\bph^2}
\def\ex{\BM{x}}
\def\exs{\ex^2}
\def\xdot{\dot{\!\ex}}
\def\y{\BM{y}}
\def\ys{\y^2}
\def\ydot{\dot{\!\y}}
\def\pat{\pa_t}
\def\pax{\pa_x}

\renewcommand{\theequation}{\arabic{equation}}


\itp{97}{102}\\

\vspace*{.3in}
\begin{center}
 \large{\bf Non-Hermitean De-Localization: \\ Multiple Scattering and Bounds}
\normalsize

\vspace{36pt}
{\bf E. Br\'ezin$^{a}$ and A. Zee$^{b}$}
\end{center}
\vskip 2mm
\begin{center}{$^{a)}$ Laboratoire de Physique
Th\'eorique,{\footnote{ Unit\'e
propre du Centre National de la Recherche Scientifique,
Associ\'ee \`a l'Ecole Normale Sup\'erieure et \`a
l'Universit\'e de
Paris-Sud}} \'Ecole Normale Sup\'erieure}\\
{24 rue Lhomond, 75231 Paris Cedex 05, France
}\\
{}
 {$^{b)}$~Institute for Theoretical Physics,}\\ {University of California, Santa Barbara, CA 93106, USA}
\vspace{.6cm}

\end{center}

\begin{minipage}{5.3in}
{\abstract~~~~~We study localization and delocalization in a class of non-hermitean Hamiltonians inspired by the problem of vortex pinning in superconductors. We show how to take into account multiple scattering. We also obtain some bounds on the complex energy spectrum.}
\end{minipage}

\vspace{48pt}
\vfill
\pagebreak

\setcounter{page}{1}

\section{Introduction}

Recently, there has been considerable interest in non-hermitean random matrix theory \cite{recent} as applied to a number of interesting physical situations. A particularly interesting issue is that of the localization-delocalization transition in non-hermitean random Hamiltonians. The earliest work to address this issue appears to be that of Miller and Wang \cite{mw} in their study of a problem in fluid flow. More recently, Hatano and Nelson \cite{hatano} have mapped the problem of the vortex line pinning in superconductors to a problem involving a non-hermitean random Hamiltonian. The prototypical Hamiltonian 
\beq\label{h}
H=H_0+W
\eeq
is the sum of 
a deterministic non-hermitean hopping term 
\beq\label{h0}
H_{0 ij} ={t\over 2}\left( e^h~\delta_{i+1, j} + e^{-h}~\delta_{i, j+1}\right)\quad , \quad i,j = 1, \cdots N
\eeq
(with the obvious periodic identification $i+N\equiv i$ of site indices) describing the drift of a vortex line being driven by an external current and 
the hermitean random potential term  
\beq\label{W}
W_{ij} = w_i~\delta_{i, j}\,.
\eeq
describing the pinning of the vortex line by impurities. Here the real numbers $w_i$ are drawn independently from some probability distribution $P\left(w\right)$ (which we will henceforth take to be even for simplicity.)
The number of sites $N$ is understood to be tending to infinity. We will also tak $t$ and $h$ to be positive for definiteness. This problem has been studied by a number of authors \cite{efetov,feinzee,zahed, beenakker, zee}
following Hatano and Nelson. Note that the Hamiltonian not only breaks 
parity as expected but is non-hermitean, and thus has complex eigenvalues. It 
is represented by a real non-symmetric matrix, with the reality implying 
that if $E$ is an eigenvalue, then $E^*$ is also an eigenvalue. In this paper, we extend and generalize the analysis in \cite{zee}, which we will refer to extensively as FZ. We will review some of the salient points in FZ briefly; for further details, the reader is referred to FZ and to the literature.

With no impurities ($w_i=0$) the Hamiltonian is immediately solvable by
Bloch's theorem with the eigenvalues
\beq\label{spectrum}
E_n = t~{\rm cos}~( {2\pi n\over N} - ih)\,, \quad (n = 0, 1, \cdots, N-1)\,, 
\eeq
tracing out an ellipse. The corresponding wave functions $\psi^{(n)}_j\sim {\rm exp}~2\pi i n j/N$ are obviously extended. 

With impurities present, ``wings" emerge out of the two ends of the ellipse. (See, for example, figure 1 of FZ. See also (\ref{arcs}, \ref{rhorealN}) below.) Some eigenvalues are now real. Evidently, the ``forks" where the two wings emerge out of
the ellipse represent a non-perturbative effect, and cannot be obtained by
treating the impurities perturbatively.

A simple argument shows that all localized eigenstates of (\ref{h}) correspond to the real eigenvalues on the wings. Conversely, the states
corresponding to complex eigenvalues are extended, that is, delocalized. In this sense, non-hermitean localization theory is, remarkably enough, simpler than hermitean localization theory.

One way to think about this problem is to imagine starting with $H_0$ and then increasing the randomness (measured by a parameter $\gam$, say) and ask if there is a critical strength of randomness $\gam_{c1}$ at which localized states first appear. There may also be another critical strength $\gam_c$ at which all the extended states become localized. Obviously another way of thinking about the problem is to start with the hermitean random problem $(h=0)$ in which according to Anderson and collaborators \cite{anderson} all states are localized. We then increase the non-hermiticity and ask about the critical strength $h_c$ of the non-hermiticity at which extended states first appear. Similarly, we can ask, for given $h$ and $\gam$, the critical hopping amplitude $t_c$ needed for the states to become de-localized.

In $FZ$, the emphasis is to understand analytically the essential physics involved, including the non-perturbative emergence of the two ``wings'' along the real axis. While some general analytic statements can be made, exact results are obtained in FZ only in two simplifying limits: the single impurity limit and the one way limit. 

In the single impurity limit, the $N$ random impurities in (\ref{W}) is replaced by a single impurity; in other words,

\begin{equation}\label{w2}
W_{ij} = w\delta_{i,1}\delta_{j,1}
\end{equation}
It was argued in $FZ$ that this simplification, though drastic, captures some of the essential physics.

In the one-way limit, the parameters in (\ref{h0}) are allowed to tend to the (maximally non-hermitean)  
limit $h\rightarrow\infty$ and $t\rightarrow 0$ such that 
\beq\label{thlim}
t~e^h\rightarrow 2
\eeq
The spectrum (\ref{spectrum}) changes into 
\beq\label{circle}
E_n = {\rm exp}~{2\pi i n\over N}\,, \quad (n = 0, 1, \cdots, N-1)\,,
\eeq
and the ellipse associated with (\ref{spectrum}) expands into the unit 
circle. 

As is standard, we are to study the Green's function

\beq\label{green}
G(z) = {1 \over z-H}
\eeq
The averaged density of eigenvalues, defined by 

\beq\label{master}
\rho(x,y) ~\equiv~ \langle {1\over N} \sum_i \delta(x-{\rm Re} E_i)~ \delta (y- {\rm Im}
E_i)\rangle \,,
\eeq
is then obtained as

\beq\label{master2}
\rho\left(x,y\right) ~=~ {1 \over \pi} {\partial \over \partial z^\ast} \overline{G\left(z\right)}
\eeq
with $z=x+iy$. Here we define

\beq
\overline{G\left(z\right)} \equiv \lag \frac{1}{N}~ {\rm tr}~ G \left(z\right) \rag
\eeq
and $< \ldots >$ denotes, as usual, averaging with respect to the probability ensemble from which $H$ is drawn. A small subtlety is that while $G(z)$ is ostensibly a function of $z$ only, $ \lag G(z) \rag$ depends on both $z$ and $z^{\ast}$. A careful discussion is given in \cite{feinzee}.

It is also useful to define the Green's function $G_0(z) = {1 \over z-H_0}$ in the absence of impurities and the corresponding $\overline{G_0(z)} \equiv {1 \over N} ~{\rm tr} ~ {1 \over z - H_0}$. For the spectrum in (\ref{spectrum}), we can use Cauchy's integration to evaluate

\beqra\label{speceval}
\overline{G_0(z)} ~&=~ &\int ~ {d \theta \over 2 \pi} ~ {1 \over z - t~ {\rm cos} (\theta - ih)} \nonumber \\
&= ~ & -{2 \over t} ~ \oint ~ {d \upsilon \over 2 \pi i} ~ {1 \over (\upsilon - \upsilon_+)(\upsilon- \upsilon_-)}
\eeqra
where we have changed variable to $\upsilon ~=~ e^{i \theta + h}$ and $\upsilon_\pm ~=~ {z \over t}~ \pm ~ \sqrt{{z^2 \over t^2} - 1}$. We obtain

\beq\label{john}
\overline{G_0(z)} ~=~ {1 \over \sqrt{z^2-t^2}} ~ \theta \left( \mid z + \sqrt{z^2 - t^2} ~ \mid - te^{h} \right)
\eeq
(There is ostensibly another term involving $\theta( \mid z ~-~ \sqrt{z^2 - t^2} \mid~-~ te^h)$. To show that this term can be dropped and that the boundary defined in the step function in (\ref{john}) is just the ellipse in (\ref{spectrum}), write $z ~=~ t~{\rm cos}(u_1 -iu_2) ~=~ t({\rm cos}~u_1 ~{\rm cosh} ~u_2 ~+i ~ {\rm sin}~u_1~ {\rm sinh} ~u_2)$ with $0\leq u_1 ~< ~ 2\pi$ and $u_2 ~\geq~ 0$. To cover the complex plane, we can restrict $u_2$ to be non-negative. We define $\sqrt{z^2 ~-~ t^2} ~=~ t({\rm cos}~u_1 ~{\rm sinh} ~u_2 ~+i ~ {\rm sin}~u_1~ {\rm cosh} ~u_2)$ so that when $z$ is in the upper half plane $\sqrt{z^2 ~-~ t^2}$ is also in the upper half plane. With these conventions, the second step function can be dropped. Then the step function in (\ref{john}) implies $u_2 ~\geq~ h$. At $u_2 ~=~ h, ~ z ~=~ x + iy$ traces out the ellipse

\beq\label{traces}
{x^2 \over ({\rm cos h} ~ h)^2} ~+~ {y^2 \over ({\rm sinh} ~ h)^2} =t^2
\eeq
in agreement with (\ref{spectrum}).
Note that this goes to the appropriate function in the hermitean limit $h ~\rightarrow~ 0$. In the one way limit, (\ref{john}) simplifies to

\beq
\overline{G_0(z)} = {1 \over z}~ \theta \left( \mid z \mid -1\right)
\eeq

In FZ, the following probability distributions were studied: the sign distribution

\beq\label{pw1}
P(w) = {1\over 2} [\delta (w-r) + \delta (w+r)]
\eeq
with some scale $r$, the box distribution 
\beq\label{box}
P(w) = {1\over 2V}~\theta (V^2-w^2)
\eeq
obtained by ``smearing" the sign distribution, and the Cauchy distribution
\beq\label{Cauchy}
P(w) = {\gam\over \pi} {1\over w^2+\gam^2}\,,
\eeq
with its long tails extending to infinity. 

We might also wish to consider the effect of diluting the impurities by setting randomly some fraction $d$ of the $w_k$'s to zero. In other words, given $P(w)$ we can consider

\beq\label{dil}
P_d(w) ~=~ d~ \delta (w) ~+~ (1-d) ~ P(w)
\eeq
with $0 \leq d \leq 1$. As $d$ increases, there should be more extended states.

The problem of determining the spectrum of $H$ with the Cauchy distribution was solved explicitly in FZ in the one-way limit. The density of eigenvalues is made of two arcs and two wings. (For completeness, we will derive a generalization of this result later in this paper.) 
The complex eigenvalues form the union of two arcs of a circle of radius one. More precisely, for $y\ne0$,

\beq\label{arcs}
\rho(x,y) = \int {\delta \theta \over 2\pi} \delta^{(2)} (z +i\gam ~{\rm sgn~(Im}~z)- e^{i\theta})
\eeq
The upper arc is that part of a semicircular arc of
a unit circle (centered at the origin) that remains in the upper half plane
after being pushed a distance $\gam$ downward along the imaginary axis. (The lower
arc is of course the mirror image of the upper arc.) Each of these arcs is thus of length ${\rm arc~cos}~(2\gam^2-1)<\pi$ and carries $n_{{\rm arc}} = (N/2\pi)~{\rm arc~cos}~(2\gam^2-1)~<~(N/2)$ eigenvalues. This means that the arcs exist only as 
long as $\gam < \gam_c =1$.  

The rest of the eigenvalues, which do not have space to live on the arcs, must have snapped onto the real axis and formed ``wings". The eigenvalue density along the real axis is given by 

\beq\label{rhorealN}
\rho_{{\rm wing}}(x,y) =  {\gam\over \pi}~{1\over x^2+\gam^2}~\theta (x^2+\gam^2-1)~\delta (y)
\eeq
We thus see that for $\gam~<~\gam_c = 1$, there are two wings that bifurcate from the arcs at $x=\pm x_c = \pm \sqrt{1-\gam^2}$. Integrating over (\ref{rhorealN}) we find that the fraction of eigenvalues that reside in 
the wings is $(n_{\rm wings}/N) = 1-(2/\pi)~{\rm arc~cos}~\gam = 1-(1/\pi)~{\rm arc~cos}~(2\gam^2-1)$, which together with the fraction of eigenvalues that reside
in the arcs, sum up exactly to $1$. As $\gam$ tends to $\gam_c$,
$x_c$ becomes smaller, and vanishes at $\gam=\gam_c$. At this point the two wings touch at the origin and the two arcs disappear completely. All the eigenvalues have snapped onto the real axis.

Note that in this problem the critical coupling $\gam_{c1}$ defined earlier is equal to $0^+$. Localized states appear no matter how small $\gam$ may be as long as it is non-vanishing. (Since we have already gone to the maximally non-hermitean limit we cannot ask about $h_c$.)

\pagebreak

\section{Multiple Scattering and ``Two Way" Hopping}

In this section we go beyond FZ and discuss what we can obtain analytically without taking the ``one way" limit in (\ref{thlim}). With the particle now able to hop both ways, the calculation of $G(z)$, which as usual can be expanded formally as a series:

\beq
G(z) ~=~ G_0(z) + G_0(z) W G_0(z) + \ldots ,,
\eeq
now involves the combinatorial problem of summing over all possible histories in which the particle, starting from some intitial site $i$, propagates to the site $j$, scatters off the impurity there, then propagates to the site $k$, scatters off the impurity there, and so on, until it arrives at some final site $f$. This is precisely what is expressed by (\ref{green}) which we can rewrite as

\beq\label{convince}
G~=~ \left(\frac{1}{1- \sum\limits_k w_k G_0 P_k}\right)G_0
\eeq
Here we have defined the projection operator
\beq
P_k ~=~ \mid k \rag \lag k \mid
\eeq

In this section we will not commit to any specific form of $H_0$. We will merely assume that $H_0$ is translation invariant.
One can average (\ref{convince}) with one's favorite $P(w)$. In practice, however, it is more useful to carry out some partial summation over the effects of repeated scattering on a specific impurity. Let us define

\beqra\label{summation}
\upsilon_k &\equiv &{w_k \over 1-w_k \lag k \mid G_0 \mid k \rag} \nonumber \\
&= &w_k + w_k^2 \lag k \mid G_0 \mid k \rag + w_k^3 \lag k \mid G_0\mid k \rag^2 ~+\cdots
\eeqra
and

\beq\label{summation2}
g_k \equiv \upsilon_k G_0 P_k G_0
\eeq
By translation invariance, $\lag k \mid G_0 \mid k \rag ~=~ \overline{G_0(z)}$. The matrix $(g_k)_{ij}$ describes the histories consisting of propagation from some initial site $i$ to the site $k$, scattering off the impurity at site $k$, once, twice, $\ldots$, any number of times, and finally propagating to some final site $j$. In other words, only scattering off the impurity at site $k$ is allowed. Next we define

\beq\label{summation3}
G_k = g_k + g_k G_0^{-1} \sum_{l\ne k} G_l
\eeq 
Then the desired Green's function is given by 

\beq
G ~=~ G_0 + \sum_k G_k
\eeq
Putting these equations together, we obtain

\beq\label{together}
G~=~ \left[1-\sum_k\left(\frac{\upsilon_k G_0 P_k}{1+\upsilon_k G_0 P_k} \right) \right]^{-1} ~ G_0
\eeq
It is straightforward to expand this expression and show that it leads to (\ref{convince}), of course. This expression, while equivalent to (\ref{convince}), is however more suitable for averaging.

For any arbitrary $P(w)$ it is presumably not possible to average (\ref{together}) explicitly and obtain $\lag G(z) \rag$ in closed form, not any more than it is possible to obtain $\lag G(z)\rag$ in closed form for the hermitean problem. Indeed, (\ref{together}) was derived by only assuming translation invariance for $H_0$, whether hermitean or non-hermitean.  

It is, however, always possible to expand (\ref{together}) to any desired powers of $\upsilon$, and average. To write down the result, it is best to define Dyson's irreducible self energy $\sum(z)$ by the standard formula

\beq\label{dyson}
\lag G\left(z\right) \rag ~=~ \frac{1}{G_0^{-1}\left(z\right)-\sum \left(z\right)}
\eeq
To third order in $\upsilon$ we obtain

\beq\label{third}
{\textstyle \sum} \left(z\right) ~=~ \frac{\lag v \rag}{1 + \lag v \rag \overline{G}_o} + \lag v \rag \left[ \lag \upsilon^2 \rag - \lag \upsilon \rag^2 \right] \left( \overline{G_o^2} - \overline{G}_o^2 \right) ~+~ O\left(\upsilon^4 \right)
\eeq
Here we have defined
\beq\label{susan}
\lag \upsilon \rag ~\equiv~ \lag \upsilon_k \rag ~=~ \lag {w \over 1-w \overline{G}_0} \rag
\eeq
and

\beq
\overline{G_0} ~=~ \lag k \mid G_0 \mid k \rag ~=~ {1 \over N} {\rm tr} G_0
\eeq
and

\beq
\overline{G_0^2} ~=~ \lag k \mid G_0^2 \mid k \rag ~=~ {1 \over N} {\rm tr} G_0^2 ~=~ -{\partial \over \partial z} \overline{G_0}
\eeq
For $H_0$ in (\ref{h}), we can read off $\overline{G_0}$ and hence $\overline{G_0^2}$ from (\ref{john}). While $\sum(z)$ is in general a matrix, it is merely a number to this order. Thus, to this order we obtain a remarkably simple result

\beq\label{bob}
\overline{G(z)} ~=~ \overline{G_0 (z-{\textstyle \sum}(z))}
\eeq
with $\sum(z)$ given in (\ref{third}). Thus,  the reader can simply compute $\lag \upsilon \rag$ with his or her favorite $P(w)$ and then apply (\ref{master2}) to (\ref{bob}) to obtain $\rho(x,y)$.

We can consider the effect of dilution. Let $<\ldots>_d$ denotes averaging with respect to $P_d(w)$ defined in (\ref{dil}). Then $\lag \upsilon^p \rag_d ~=~ (1-d)~ \lag \upsilon^p \rag$. Thus $\Sigma_d(z)$ is obtained to third order in $\upsilon$ from the $\Sigma(z)$ in (\ref{third}) by replacing the moments $\lag \upsilon^p \rag$ with $\lag \upsilon^p \rag_d$. We can thus in principle obtain $\rho(x,y)$ for any value of the dilution. In particular, for $d \rightarrow 1$, we obtain the simple but uninteresting result

\beq\label{dist}
\overline{G(z)} ~=~ \overline{G_0 (z ~-~ (1-d) ~\lag \upsilon(z) \rag )}
\eeq
The ellipse in (\ref{traces}) is distorted slightly.

For the Cauchy distribution, things simplify enormously. Recall that if the random variable $x$ obeys a Cauchy distribution, then $(x ~+~ {\rm constant})$ and $x^{-1}$ also obey Cauchy distributions. Thus $\upsilon_k$ obeys a Cauchy distribution and $\lag \upsilon^p \rag ~=~ \lag\upsilon \rag^p$ for any integer power $p$. (Note that in contrast to the moments $\lag \upsilon^p \rag$  the moments $\lag w^p \rag$ do not exist. See (\ref{susan}).) With $\lag \upsilon^2 \rag = \lag \upsilon \rag^2$, the expression in (\ref{third}) suggests that $\sum (z)$ is given to all orders simply by $\sum (z) ~=~ {\lag \upsilon \rag \over 1 + \lag \upsilon \rag ~ \overline{G_0} }$ . We will now sketch a proof that this is indeed the case.

Indeed with the Cauchy distribution the hermitean version $\left( h \rightarrow 0 \right)$ of our problem has long been known as the exactly solvable Lloyd model \cite{zinman}. The key to the solvability is the fact that the averaging integrals over the $w_k$'s can all be done using Cauchy's theorem. For $z$ in the upper half plane, the integration contours in $w$ can be closed in the lower half plane and $w$ is effectively set equal to $-i \gamma$, and vice versa if $z$ is in the lower half plane. We must now show that this procedure, proved in the literature for hermitean Hamiltonians, continues to hold for non-hermitean Hamiltonians. Imagine expanding (\ref{together}) in a series in the $\upsilon_k$'s. Upon averaging, the $\upsilon_k$'s are effectively replaced by $\lag \upsilon \rag$. Using

\beq
{1 \over 1 ~+~ \lag v \rag ~ G_0 P_k} ~=~ 1 ~ - ~ {\lag \upsilon \rag G_0P_k \over 1 ~+~ \lag \upsilon \rag ~ \overline{G_0}}
\eeq
we obtain from (\ref{together}) that indeed

\beq
{\textstyle \sum} (z) ~=~ {\lag \upsilon \rag \over 1 + \lag \upsilon \rag ~ \overline{G_0} }
\eeq
We see from (\ref{susan}) that $\sum(z)$ is equal to $\mp i \gamma$ in the upper and lower half plane respectively. We thus obtain from (\ref{dyson})

\beq\label{L}
\overline{G(z)} ~=~ \overline{G_0 \left( z + i \gamma \right)} ~ \theta \left( {\rm Im} ~z \right) ~+~ \overline{G_0 \left(z-i \gamma \right)} ~ \theta\left({\rm -Im}~ z \right)
\eeq
The density of eigenvalues again consists of two arcs and two wings. When ${\partial \over \partial z^{\ast}}$ in (\ref{master2}) acts on $\overline{G_0}$ in (\ref{L}) we obtain

\beq\label{M}
\rho(x,y) ~=~ \rho_0(x,y + \gamma) ~ \theta(y) ~+~ \rho_0(x, y- \gamma) ~ \theta(-y)
\eeq
In other words, the two arcs of the ellipse are pushed towards the real axis by a distance $\gamma$. When ${\partial \over \partial z^{\ast}}$ in (\ref{master2}) acts on the step functions in (\ref{L}) we obtain the density on the two wings

\beq\label{N}
\rho_{wing}(x,y) ~=~ {1 \over \sqrt{2} \pi} ~ \delta (y) ~ \theta (x^2-x^2_{min} (\gamma)) ~ {\sqrt{\gamma^2-x^2+t^2 ~+~ B(x,\gamma,t)} \over B(x, \gamma, t)}
\eeq
where we have defined for convenience
\beq
B(x, \gamma, t) ~\equiv~ \sqrt{(x^2 + \gamma^2)^2 ~+~ 2t^2(\gamma^2-x^2) ~+~ t^4}
\eeq
and

\beq\label{conv}
x_{min}(\gamma) ~\equiv~ (\sqrt{(t ~{\rm sinh}~ h)^2 ~-~ \gamma^2})~ {\rm tanh}~ h
\eeq
The critical value of $\gamma$ at which the arcs disappear is given by 

\beq\label{critical}
\gamma_c ~=~ t ~{\rm sinh}~ h
\eeq
All states are now localized. Evidently, the geometrical construction described after (\ref{arcs}) still works. In the one way limit, all these quantities reduce appropriately and $\rho(x,y)$ tends to (\ref{arcs}) and (\ref{rhorealN}).

In this model, $\gam_{c1}$ is again $0^+$. The wings disappear as $\gam \rightarrow 0$. We see that for insufficient non-hermiticity, $h~<~h_c$ where

\beq
h_c ~=~ {\rm sinh}^{-1} ~ ({\gam \over t}) \,,
\eeq
all states are localized. Note that in accordance with Anderson et al \cite{anderson}, for any finite non-zero $\gam$, no matter how small, there are localized states for $h$ small enough. Trivially, we can express (\ref{critical}) in yet another way, by saying that for a given non-hermiticity $h$ and site randomness $\gam$, we need the hopping to exceed a critical strength

\beq\label{critstrength}
t_c ~=~ {\gam \over {\rm sinh} ~h}
\eeq
before states become delocalized.
\pagebreak
\section{Bounds in the One Way Limit}

In this section we revisit the one way limit and extend the analysis in FZ. In the one way limit, the determinant of $z-H$ simplifies drastically to (we take $N$ to be even for definiteness)
\beq\label{det}
{\rm det}~(z-H) = \left(\prod_{k=1}^N (z-w_i)\right) -1\,
\eeq
(For arbitrary $t$ and $h$ the corresponding formula is considerably more complicated.) Note that (\ref{det}) is completely symmetric in the $\{w_i\}$, and thus, for a given set of site energies it is independent of the way the impurities are 
arranged along the ring. Differentiating the logarithim of ${\rm det} (z-H)$ with respect to $z$ we obtain formally

\beq\label{logdet}
\frac{1}{N}~{\rm tr} ~\frac{1}{z-H} ~=~ \left( \frac{1}{N}~\sum_j \frac{1}{z-w_j} \right) \sum_{l=0}^{\infty} \left(\prod_i \frac{1}{z-w_i} \right)^{l}
\end{equation}
Defining

\begin{equation}
g\left( z \right) \equiv \langle \left( \frac{1}{z-w} \right) \rangle
\end{equation}
and

\begin{equation}
g_l \left(z \right) \equiv \langle \left(\frac{1}{z-w} \right)^l \rangle = \frac{\left(-1 \right)^{l-1}}{\left(l-1\right)!} \left(\frac{d}{dz} \right)^{l-1} g\left(z \right)
\end{equation}
we obtain

\beq\label{obtain}
\overline{G(z)} ~=~ \sum_{l=0}^{\infty} g_{l+1} \left(z \right) \left(g_{l}\left(z \right) \right)^{N-1}
\eeq
We are to evaluate this in the large $N$ limit and then compute the density of eigenvalues by $\rho\left(x,y \right)=\frac{1}{\pi}\frac{\partial}{\partial z^{\ast}} \overline{G} \left(z \right)$
For an arbitrary $P(w)$ it is non-trivial to evaluate $g_{l}\left(z \right)$ and to carry out the sum in (\ref{obtain}).

In FZ it was noted that in the case of the Cauchy distribution $P(w)$ this program can be carried out. It is worthwhile, so that we can see how the expression in (\ref{obtain}) works, to use (\ref{obtain}) to give a somewhat more compact derivation of (\ref{arcs}) and (\ref{rhorealN}) here than in FZ. The key is that in the Cauchy case $g_l(z)$ is equal to $g(z)^l$, a simple power of 

\beq\label{simple}
g(z) ~=~ {1 \over z + i\gamma ~ {\rm sign} \left[ {\rm Im}~ z \right]}
\eeq
a result which can be obtained easily by using Cauchy's formula for contour integration. The sum in (\ref{obtain}) is now immediate, giving $\overline{G(z)} = \frac{g(z)}{1-g(z)^N}$ (where we neglect the difference between $N$ and $N-1$). Thus, for $\mid g(z) \mid < 1$, we have $\overline{G(z)} = g(z)$, and for $\mid g(z) \mid > 1$, we have $\overline{G(z)} = 0$. We obtain

\beqra
\overline{G(z)} &= &g(z) \theta \left(1- \mid g(z) \mid \right) \nonumber \\
&= & \frac{1}{x+i\left(y+ \gamma\right)} \theta \left(x^2 + \left(y+ \gamma \right)^2 -1\right) \theta \left(y\right) \nonumber \\
&+ & \frac{1}{x+i\left(y- \gamma\right)} \theta \left(x^2 + \left(y -\gamma \right)^2 -1\right) \theta \left(-y\right) 
\eeqra
Differentiating according to (\ref{master2}) we obtained the density of eigenvalues in (\ref{arcs}) and (\ref{rhorealN}) with its arcs and wings.

While we are not able to evaluate the sum in (\ref{obtain}) for $P(w)$ other than the Cauchy distribution, we can devise bounds on the domain over which the density of eigenvalues is non-zero. Indeed, the analysis given in the Cauchy case suggests how to proceed.  We will assume that $P(w)$ has bounded support; an example is the box distribution. We will always take $P(w)$ to be even for simplicity. Note that $g_l(z) = g_l(-z)$ then and $\rho (x,y) ~=~ \rho (-x,-y) ~=~ \rho (x,-y) ~=~ \rho (-x,y)$. This allows us to restrict the subsequent analysis to the first quadrant ($x>0$ and $y>0$).

We begin by noting that if $\mid g_{l^{\ast}} \mid < 1$, then the $l^{\ast}$ term in the sum in (\ref{obtain}) can be dropped. This inequality defines a domain $D\left(x,y;l^{\ast}, \{r\}\right)$ in the complex plane. (Here $\{r\}$ denotes the set of parameters in $P(w)$.) The intersection of all such domains $D\left(x,y;l^{\ast},\{r\}\right)$, as $l^{\ast}$ ranges over all integers from $l^{\ast} = 1$ to infinity, defines a domain $D$ over which the density of eigenvalues $\rho\left(x,y\right)$ vanishes and which may of course  be considerably smaller than the actual domain over which $\rho\left(x,y\right)$ vanishes. Note the $l=0$ term in (\ref{obtain}) can of course never be dropped. Ostensibly, it contributes to $\overline{G(z)}$ a term $g(z)$ and hence, according to (\ref{master2}), to $\rho(x,y)$ the term $\lag \delta(x-w)~ \delta(y) \rag ~=~ P(x) ~ \delta(y)$. For $P(w)$ of bounded support, this represents a finite line segment and as long as it does not protrude into $D$ it is not relevant to our search for a bound on $D$. Whether or not such a line segment actually appears in $\rho(x,y)$ is an issue upon which we will remark later. 

Given the preceding observation, it is easy to obtain some bounds. Our analysis to follow will be rather informal; however, we believe that all the bounds we obtain could be made mathematically rigorous.

As a simple first example of a bound, consider

\begin{equation}
\begin{array}{lll}
\mid g_l(iy) \mid &\leq & \int dw P(w) \mid \frac{1}{iy-w} \mid^{l}\\
&\leq & \int dw P(w) \left(\frac{1}{y}\right)^{l} = \frac{1}{y^l}
\end{array}
\end{equation}
which is less than 1 for all positive integers $l$ if $\mid y \mid >1$. Thus, we conclude that for $\mid y \mid >1$ the density of eigenvalues vanishes. This indeed holds true for all the examples worked out in FZ. 

As a second example, we note that 

\beqra\label{bound}
\mid g_{l}\left(z\right) \mid &= &\int dw P(w) \mid \frac{1}{x+iy-w} \mid^l\\
&= &\int dw P(w) \mid \left(\frac{1}{(x-w)^2 + y^2}\right)^{\frac{l}{2}} \mid 
\eeqra
By definition, $x-w ~>~ x-w_{max}$, and hence $(x-w)^2 ~>~ (x-w_{max})^2$ if $x>w_{max}$. We obtain

\beq
\mid g_l(z) \mid ~<~ \left( \frac{1}{(x-w_{max})^2 +y^2}\right)^{\frac{l}{2}}
\eeq
and thus there is no density of eigenvalues if

\beq\label{noden}
(x-w_{max})^2 + y^2 ~>~ 1
\eeq
and

\beq\label{noden2}
x>w_{max}
\eeq

Combining this with the result obtained above, we conclude that the density of eigenvalues is confined to an oval-shaped region defined by $y>1$ and (\ref{noden}) and (\ref{noden2}). (In drawing a figure, it is useful to consider the cases $w_{max} > 1$ and $w_{max} < 1$ separately.) In particular, we conclude that the tips of the wings cannot extend farther than $(w_{max} +1)$. (The Cauchy distribution does not have bounded support and indeed the wings extend to infinity.)

The bound (\ref{bound}) can be improved if $P(w)$ vanishes for $-w_{min}<w<w_{min}$, in other words, if there is a hole in the middle of the distribution. The sign model defined by (\ref{pw1}) provides an example of this class. In the integral defining $g_l(z)$, we have $\left(x-w\right)^2 > \left(x+w_{min}\right)^2$ for $w<0$ and $\left(x-w\right)^2 > \left(x-w_{min}\right)^2$ for $w>0$. We find that the condition $\mid g_l(z) \mid<1$ leads to 

\beq\label{inequality}
\left(\frac{1}{\left(x+w_{min}\right)^2+y^2}\right)^{\frac{l}{2}} + \left(\frac{1}{\left(x-w_{min}\right)^2+y^2}\right)^{\frac{l}{2}} < 2
\eeq

In particular, at $x=0$ we find that $\rho(x,y)~=~0$ when $y^2 > 1 - w_{min}^2$. Thus, for $w_{min}^2>1$ the entire $y$ axis is free from eigenvalues, and the eigenvalue distribution has split into, by symmetry, at least two separate ``blobs." This was found to be indeed the case in the sign model in FZ with the critical value $r_c=1$. (See figure (2c) in \cite{zee}).

The inequality (\ref{inequality}) can also be analyzed for $x \ne 0$. By symmetry we can take $x>0$. Define $s^2=\left(x+w_{min}\right)^2+y^2$. Then we can rewrite (\ref{inequality}) as

\begin{equation}
2s^l > 1 + \frac{s^l}{\left(s^2-4w_{min} x \right)^{\frac{l}{2}}}
\end{equation}
The inequality is satisfied for $(s^2-4w_{min} x) ~>~ 1$ and thus for

\begin{equation}
y>\sqrt{1-\left(x-w_{min}\right)^2} ~>~ \sqrt{1 - w_{min}^2}
\end{equation}
There is a similar inequality for $x<0$ obtained by reflection.
\pagebreak

\section{Sign Model and Expulsion of Eigenvalues}

We now extend the analysis given in FZ of the sign model in the one way limit. For this model, using (\ref{det}) we obtain

\beq\label{therm1}
\langle {\rm tr~log}~(z-H)\rangle ~=~ 2^{-N} \sum_{n=0}^N {N \choose n}{\rm  log}~\left[(z-r)^{n} (z+r)^{N-n} -1\right]\,.
\eeq
This expression has the simple physical interpretation that for a specific realization of $H$, $n$ of the impurity potentials have the value $+r$ while the others have the value $-r$. Define $n ~=~ {N \over 2}~ (1+\sigma)$. Then for this specific realization, the eigenvalues are given by

\beq\label{realize}
(z^2-r^2) ~ ( {z-r \over z+r} )^{\sigma} ~=~ e^{i {4 \pi j \over N}} \rightarrow e^{i \theta}
\eeq
Here $j=1, \ldots, N$, and the $N \rightarrow \infty$ limit is indicated with $\theta$ running between $0$ and $2 \pi$. Thus, the density of eigenvalues $\rho (x,y)$ spans a one dimensional curve for each realization. For finite $N$, the density of eigenvalues is formed by superposing these curves (See figure (2) in FZ.) In the limit $N\rightarrow\infty$, the binomial coefficient appearing in (\ref{therm1}) is sharply peaked around $n\sim N/2$, and can be approximated by $e^{- {N \sigma^{2} \over 2}}$. Thus, the range in $\sigma$ vanishes like ${1 \over \sqrt{N}}$. (This is consistent with the width of the apparently two-dimensional regions obtained by diagonalizing finite $N$ Hamiltonians in figure (2) in FZ.)

Thus, we see that the original unit circle spectrum of the 
deterministic $H_0$ is distorted by randomness into the curve 

\beq\label{curve}
z^2=r^2 + e^{i\theta}\,,\quad 0\leq\theta<2\pi
\eeq
in the complex $z$ plane. Clearly, $r_c=1$ is a critical value of $r$. For $r<1$, the curve (\ref{curve}) is connected, enclosing a region free of energy eigenvalues. For $r>1$ it breaks into two disjoint symmetric lobes that are located to the right and to the left of the imaginary axis. 
We mention in passing that if we substitute $z \rightarrow iy$ in (\ref{realize}) we obtain the solution

\beq
y^2 ~=~ 1-r^2
\eeq
independent of $\sigma$ even for finite $N$. This is consistent with figure (2) in FZ.

We mention that our result for the sign model satisfies the bounds we just derived, of course. For instance, the curve (\ref{curve}) intersects the positive real axis at $E_{max} ~=~ \sqrt{r^2 + 1}$ (and at $E_{min} ~=~ \sqrt{r^2 - 1}$ if $r>1$.) The bound that the tip of the wing cannot extend beyond ($w_{max}~+~1$) is certainly satisfied since $(r+1) ~>~ \sqrt{r^2 + 1}$ for $r$ positive. Note also that by our discussion there are no wings in the sign model. This can also be argued heuristically from (\ref{obtain}) since term by term there is no cut in $g_l (z)$. Thus, the critical strength of randomness at which localized states first appear (called $\gam_{c1}$ in the Introduction) is effectively infinite.

One feature exemplified by our result for the Cauchy model and the sign model is the presence of a ``hole" in the spectrum, that is, there are no eigenvalues near $z \sim 0$. In the one way limit we can understand this feature analytically. Rewrite (\ref{logdet}) as

\beq\label{rewrite} 
{1 \over N} ~ {\rm tr} ~ G(z) ~=~ \frac{1}{\prod\limits_i (z-w_i)-1} ~ \left[ \sum_k ~ \prod_{j\ne k} \left(z-w_j \right) \right]
\eeq
The expression in the square bracket is a polynomial. Thus, near $z \sim 0$, $G(z)$ is singular only if $\prod_iw_i = 1$. So the presence of the hole is related to the probability that $\prod_i w_i =1$. This provides another understanding of the critical transition at $r_c = 1$ in the sign model: the probability just referred to becomes ${1 \over 2}$ at that point. Our simple argument suggests that the presence of the hole is generic for any $P(w)$ which does not have any singularity at $w=1$ (such as the box distribution). See figure 1 of FZ.

Earlier, we mentioned that the $l=0$ term in (\ref{obtain}) apparently contributes a term $P(x) ~ \delta(y)$ to $\rho (x,y)$, representing a finite line segment for $P(w)$'s with bounded support, such as in the box model, or representing two points, such as in the sign model. From the argument just given and from the explicit result (\ref{curve}) this contribution does not seem to be there. We believe that this expulsion of eigenvalues from the region around $z \sim 0$ is a generic phenomenon.

In order to have a clear understanding of this phenomenon, it is perhaps better to give a simple explicit demonstration, rather thatn a heavy formal proof. We consider the simplest possible situation, combining the one way limit and the single impurity limit. Schr${\rm \ddot{o}}$dinger's equation reads

\beqra
(E-w) ~ \psi_1 &= &\psi_2 \\
E~\psi_2 &= &\psi_3 \\ \nonumber
&\vdots \\
E~\psi_{N-1} &= &\psi_N \\
E~\psi_N &= &\psi_1
\eeqra
In the absence of the impurity, the spectrum is clearly given by the unit circle $\mid E \mid = 1$.

Suppose in a particular realization of the random energy $w$ we have $w>1$. We see that there is a solution $\psi_N ~=~ {1 \over w}~ \psi_1~,~ \psi_{N-1} ~=~ {1 \over w^2}~ \psi_1~,~ \ldots~,$ which decreases rapidly, so that by the time we reach the site $2$, $\psi_2$ is exponentially small. Thus, we have a real eigenvalue $E=w$ to exponential accuracy. A similar discussion can be given for $w<-1$. In contrast, in a realization in which $\mid w \mid < 1$, the solution just described becomes unbounded and must be excluded.

As we approach the circle moving along either of the two wings (which have spectral weight ${1 \over N}$), in other words, for $E$ real and tending towards $\pm 1$, the localization length diverges as

\beq\label{diverge}
L(E) ~\sim~ {1 \over {\rm log}~\mid E \mid}
\eeq

It is tempting to conjecture that in general, if $E_c$ is the point at which the (right) wing attaches to the complex spectrum (for example, $x_{min}(\gam)$ in (\ref{conv})), the localization length diverges as

\beq
L(E) ~\sim~ {1 \over {\rm log}~ {E \over E_c}}
\eeq
as $E \rightarrow E_c$.

Finally, we mention for future work that given (\ref{logdet}) we can write down easily in the one way limit a formal expression for the correlation function

\beq\label{correlation}
\overline{G(z,z^{^{\prime}})} ~ \equiv ~ \lag {1 \over N}~ {\rm tr}~ {1 \over z-H}~ {1 \over N} {\rm tr}~ {1 \over z^{^{\prime}} -H} \rag
\eeq
much studied in recent years in the context of random matrix theory. Define

\beqra
h(z,z^{^{\prime}}) &\equiv &\lag ( {1 \over z-w} ) ({1 \over z^{^{\prime}}-w} ) \rag \nonumber \\
&= & (-) ~~ {g(z) - g(z^{^{\prime}}) \over z - z^{^{\prime}}}
\eeqra
and

\beq
h_{ll^{^{\prime}}} (z,z^{^{\prime}}) ~=~ \lag ({1 \over z-w} )^l ~ ( {1 \over z^{^{\prime}} - w} )^{l^{^{\prime}}} \rag
\eeq
(Of course $h_{ll^{^{\prime}}}$ can be obtained from $h$ by differentiation.) We then obtain

\beqra
\overline{G(z,z^{^{\prime}}}) &= &{1 \over N} \sum_{l=0}^{\infty} \sum_{l^{^{\prime}}=0}^{\infty} h_{l+1,l^{^{\prime}} +1} (z,z^{^{\prime}}) (h_{ll^{^{\prime}}} (z,z^{^{\prime}}))^{N-1} \nonumber \\
&+ &(N-1) h_{l+1,l^{^{\prime}}} (z,z^{\prime}) h_{l,l^{\prime} +1} (z,z^{\prime}) \left(h_{ll^{\prime}}(z,z^{\prime}) \right)^{N-2}]
\eeqra
For an arbitrary $P(w)$ this is certainly no easier to evaluate in closed form than it is to evaluate $\overline{G(z)}$. (It is however simple to show that in the Cauchy case $\overline{G(z,z^{\prime}})$ has no connected piece for $z$ and $z^{\prime}$ on the same side of the real axis.)

\pagebreak
\section{Conclusion}

We have studied the spectrum of a non-hermitean hopping Hamiltonian, with real eigenvalues corresponding to localized states and complex eigenvalues corresponding to extended states. Going beyond an earlier work FZ, we treat multiple scattering, obtaining a general formula, which we evaluated to a certain finite order, arriving at an expression (\ref{bob}) from which one can compute the density of eigenvalues for an arbitrary probability distribution $P(w)$ of the random site energies. For the Cauchy distribution, we obtain explicit analytic results. In the so-called one way limit, while it is not possible to give explicit analytic results for an arbitrary $P(w)$, we are able to obtain bounds on the spectrum of the Hamiltonian. The spectrum of the so-called sign model was worked out explicitly and shown to satisfy the bounds derived. We also give an analytic argument for the expulsion of eigenvalues from the region around $z \sim 0$, a phenomenon we believe to be generic.

Note: While completing this paper, we noticed the paper of Goldsheid and Khoruzhenko \cite{IY} who also treated the Hamiltonian in (\ref{h}).

{\bf Acknowledgements}~~~
One of us (A. Zee) thanks the \'Ecole Normale Sup\'erieure, where part of this work was done, for its warm hopitality and support. This work was partly supported by the National Science Foundation under Grant No. PHY89-04035.

\end{document}